\pdfoutput=1

\documentclass[11pt,twoside,a4paper,cmspaper,final,collab]{cms-tdr}
\def\svnVersion{166196}\def\cmsCernNo{2011-221}\def\cmsCernDate{\today}\def\cmsMessage{Submitted to Physics Letters B}

\begin{document}\cmsNoteHeader{HIN-11-002}

\hyphenation{had-ron-i-za-tion}
\hyphenation{cal-or-i-me-ter}
\hyphenation{de-vices}
\RCS$Revision: 105421 $
\RCS$HeadURL: svn+ssh://svn.cern.ch/reps/tdr2/papers/HIN-11-002/trunk/HIN-11-002.tex $
\RCS$Id: HIN-11-002.tex 105421 2012-02-20 13:29:03Z yjlee $
\newcommand{\snn}{\ensuremath{\sqrt{s_{_\text{NN}}}}} 
\cmsNoteHeader{HIN-11-002} 
\title{Measurement of isolated photon production in pp and PbPb collisions at $\sqrt{s_\mathrm{NN}}$ = 2.76 TeV}

\date{\today}

\abstract{
Isolated photon production is measured in proton-proton and lead-lead collisions at nucleon-nucleon centre-of-mass energies of 2.76~TeV in the pseudorapidity range $|\eta|<$~1.44 and transverse energies \ET between 20 and 80\GeV with the CMS detector at the LHC. The measured \ET spectra are found to be in good agreement with next-to-leading-order perturbative QCD predictions. The ratio of PbPb to \Pp\Pp\ isolated photon \ET-differential yields, scaled by the number of incoherent nucleon-nucleon collisions, is consistent with unity for all PbPb reaction centralities.}

\hypersetup{%
pdfauthor={CMS Collaboration},%
pdftitle={Measurement of isolated photon production in pp and PbPb collisions at sqrt(sNN) = 2.76 TeV.},%
pdfsubject={CMS},%
pdfkeywords={CMS, physics, heavy-ions, photons}}

\maketitle 

\section{Introduction}

Prompt photons with high transverse energy (\ET) in hadronic collisions are produced directly from the hard scattering of two partons.
At lowest order in perturbative QCD calculations,
three partonic mechanisms produce prompt photons in hadronic collisions:
(i) quark-gluon Compton scattering $\cPq\cPg\rightarrow\Pgg\cPq$, (ii) quark-antiquark annihilation
$\cPq\cPaq\to\Pgg\cPg$, and (iii) collinear fragmentation of a final-state parton into a
photon. Prompt photons from (i) and (ii) are called ``direct''; those from
(iii) are called ``fragmentation''. Measured photon production cross sections provide a direct test of perturbative
quantum chromodynamics (pQCD)~\cite{JETPHOX}, and constrain the proton~\cite{Ichou:2010wc} and nuclear~\cite{Arleo:2011gc} parton distribution functions (PDFs).
In the case of nuclear collisions, jets are significantly suppressed~\cite{Chatrchyan:2011sx,Atlas:2010bu} but direct photons as well as W and Z bosons~\cite{Chatrchyan:2011ua,ATLAS:2010px} are unaffected by the strongly interacting medium produced in the reaction. Thus, these electroweak particles constitute particularly ``clean" probes of the initial state of the collision. In particular,
the direct comparison of production
cross sections of such probes in \Pp\Pp\ and nuclear collisions allows one to
estimate possible modifications of the
nuclear parton densities with respect to a simple incoherent
superposition of nucleon PDFs.

However, the measurement of prompt photon production is complicated by the presence of a large background
coming from the electromagnetic decays of neutral mesons (mostly $\Pgpz,\Pgh\to\Pgg\,\Pgg$)
produced in the fragmentation of hard-scattered partons. Since high-transverse-momentum (\pt) neutral mesons are produced
inside a jet, they are surrounded by significant hadronic activity from other parton fragments.
Thus, $\Pgg$ backgrounds from these decays are typically suppressed by imposing isolation
requirements on the reconstructed photon candidates. The isolation requirements also significantly suppress the fragmentation photon component, while removing very few of the photons arising from direct processes. Since the annihilation contribution is relatively small at the Large Hadron Collider (LHC), the result is an isolated photon sample dominated by quark-gluon Compton photons~\cite{Ichou:2010wc}. In heavy-ion collisions, the hard scattering that produces an isolated
photon is superimposed on the considerable activity arising from multiple parton-parton scatterings (underlying event) occurring simultaneously. A subtraction of the underlying event is therefore necessary before applying isolation criteria.

In this paper, a measurement of the isolated photon production in \Pp\Pp\ and PbPb collisions at nucleon-nucleon centre-of-mass energies $\snn=2.76\TeV$ with the Compact Muon Solenoid (CMS) detector~\cite{JINST} is reported. This constitutes the first measurement of isolated photon production in heavy-ion collisions (though inclusive single photon production has been measured previously at RHIC~\cite{Adler:2005ig} and SPS~\cite{Aggarwal:2000th} energies). Sections \ref{ref:exp} and \ref{chap:triggering} describe the detector and triggers used in the analysis, while the Monte Carlo (MC) simulation and the PbPb reaction centrality
determination are discussed in Sections~\ref{sec:MC} and \ref{section:centrality}.
The photon reconstruction and identification methods used in \Pp\Pp\ collisions follow very closely those described in the studies at $\sqrt{s}=7\TeV$~\cite{CMSppPhoton}. The improvements introduced in order to adapt the photon reconstruction and isolation
to the high-multiplicity PbPb environment are discussed in Section~\ref{photonRecoID}. The photon signal
extraction and corrections are discussed in Section~\ref{sec:signal_ext}.
The theoretical pQCD calculations from the \textsc{jetphox} program~\cite{JETPHOX} are presented in Section~\ref{sec:Theory}.
Finally, the measured isolated photon \et spectra in \Pp\Pp\ and PbPb collisions are compared to the theory and to each other in
Section~\ref{sec:results}.

\section{The CMS detector}
\label{ref:exp}

Final-state particles produced in the \Pp\Pp\ and PbPb collisions are measured and reconstructed
in the CMS detector, consisting of several sub-detector systems~\cite{JINST}.
The central tracking system comprises silicon pixel
and strip detectors that allow for the reconstruction of the trajectories of
charged particles in the pseudorapidity
range $|\eta| < 2.5$, where $\eta = -\ln [\tan(\theta/2)]$
and $\theta$ is the polar angle relative to the counterclockwise
beam direction. CMS uses a right-handed coordinate system, in which
the $z$ axis runs along the beam, the $y$ axis is directed
upwards, and the $x$ axis lies in the accelerator plane and points towards
the center of the LHC ring.
Electromagnetic (ECAL) and hadron (HCAL)
calorimeters are located outside the tracking system and provide
coverage for $|\eta| < 3$. In the central (``barrel'') pseudorapidity range
$|\eta|<1.44$ considered in this analysis, the ECAL and HCAL calorimeters are finely segmented
with a granularity of $0.0174 \times 0.0174$ and $0.087 \times 0.087$, respectively,
in $\eta$ and azimuthal angle $\phi$ (in radians). 
The calorimeters and tracking systems
are located within the 3.8 T magnetic field of the super-conducting solenoid.
In addition to the barrel and endcap detectors, CMS includes a hadron forward (HF) steel/quartz-fibre Cherenkov calorimeter, which covers the forward rapidities $3<|\eta|<5.2$ and is used to determine the degree of overlap (``centrality'') of the two colliding Pb nuclei. A set of scintillator tiles, the beam scintillator counters (BSC), is mounted on the inner side of the HF for triggering and beam-halo rejection for both \Pp\Pp\ and PbPb collisions.

\section{Data samples, triggers and event selection}
\label{chap:triggering}

\begin{figure}[hbtp]
  \begin{center}
    \includegraphics[width=0.48\textwidth]{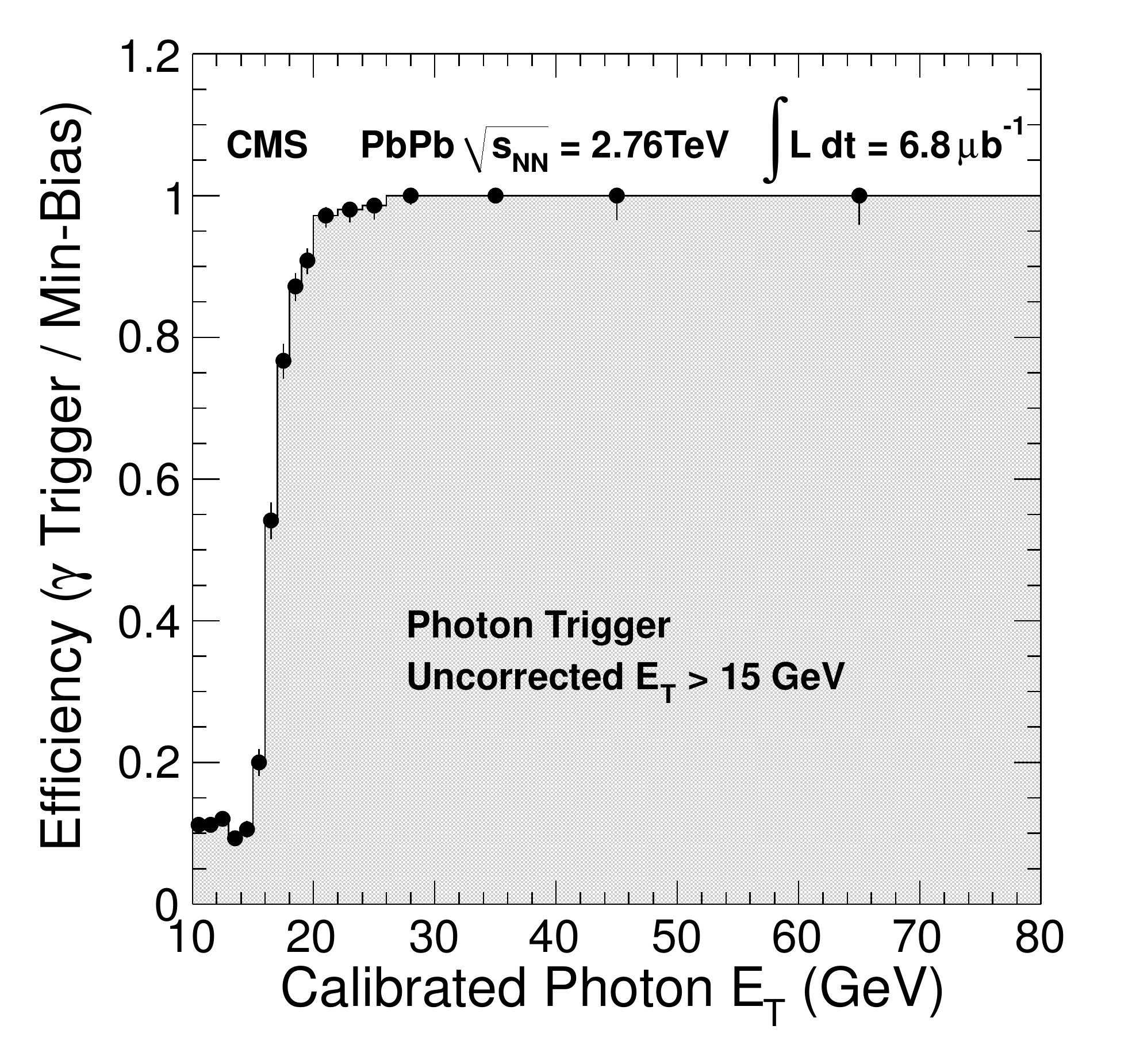}
    \caption{ Efficiency for the photon trigger as a function of the corrected photon transverse
      energy in PbPb collisions at 2.76\TeV, measured with the minimum-bias sample. Error bars represent the statistical uncertainty.}
    \label{trigger_turn_on_curve}
  \end{center}
\end{figure}

The results presented here are based on inclusive photon samples collected in \Pp\Pp\ and PbPb collisions at 2.76\TeV with minimum-bias and photon triggers. 
The total data sample corresponds to an integrated luminosity of 231\nbinv and $6.8\,\mu\mathrm{b}^{-1}$
for \Pp\Pp\ and PbPb, respectively. Note that the \Pp\Pp-equivalent luminosity of the PbPb measurement,
${\cal L}_\text{pp-equiv} = A^2\times{\cal L}_\mathrm{PbPb}=294\nbinv$ (where A=208 is the nuclear mass number for Pb), is close to that of the \Pp\Pp\ data.
For online event selection, CMS uses a
two-level trigger system: a level-1 (L1) and a high level trigger (HLT).
The trigger and event selection used for the \Pp\Pp\ analysis are described elsewhere~\cite{CMSppPhoton}. PbPb events used in
this analysis are selected by requiring a L1 electromagnetic cluster with $\ET> 5\GeV$ and an HLT photon
with $\ET > 15\GeV$, where \et values do not include offline corrections for the calorimeter energy response. The
efficiency of the photon trigger in PbPb collisions is shown in Fig.~\ref{trigger_turn_on_curve} for photon candidates with $\abs{\eta^\gamma}<1.44$. The efficiency is greater than $98\%$ for
photon candidates with corrected transverse energy $\ET^\gamma > 20\GeV$ in both \Pp\Pp\ and PbPb collisions.

In addition to the photon-triggered data sample, a minimum-bias (MB) PbPb event sample is
collected using coincidences between trigger signals from the $+z$ and $-z$ sides of either the BSC or the HF. The minimum-bias trigger and event selection efficiency in PbPb collisions is $(97\pm 3)\%$~\cite{Chatrchyan:2011sx}.

To select a pure sample of inelastic hadronic PbPb collisions, the contamination from electromagnetic (``ultra-peripheral'') collisions and non-collision beam background are removed following the prescriptions in Ref.~\cite{Chatrchyan:2011sx}. Events are preselected if they contain a reconstructed vertex made of at least two tracks with vertex $z$ position $|z|< 15$ cm and an offline HF coincidence of at least three towers with energy greater than 3\GeV on each side of the interaction point. To further suppress the beam-gas and beam-scraping events, the length of pixel clusters along the beam direction is required to be compatible with particles originating from the event vertex.

Offline selection of \Pp\Pp\ and PbPb events for further analysis requires a photon candidate, defined as described in
Section~\ref{photonRecoID}, in the pseudorapidity range $\abs{\eta^\Pgg}<1.44$ and with a corrected transverse energy $\ET^\Pgg > 20\GeV$, defining the phase space of the measurement.

\section {Monte Carlo simulation}
\label{sec:MC}
In order to study the photon selection efficiency and electron rejection in PbPb collisions,
$\Pgg$+jet, dijet, and $\PW\rightarrow \Pe\cPgn$ events are simulated using the \PYTHIA  Monte Carlo (MC) generator (version 6.422, tune D6T)~\cite{Sjostrand:2006za}, modified
to take into account the isospin of the colliding nuclei~\cite{Lokhtin:2005px}. These simulated \PYTHIA events, propagated through the CMS detector using the \GEANTfour package~\cite{geant4} to simulate the detector response, are embedded in actual MB PbPb events in order to study the effect of the
underlying event on the photon reconstruction and isolation.
The embedding is done by mixing the simulated digital information with the recorded MB PbPb data. These mixed samples (denoted ``\textsc{pythia+data}") are used for signal shape studies, and for energy and efficiency corrections.

In order to determine whether a given photon is isolated at the generator level, an isolation cone of radius $\Delta R$ = $\sqrt{(\Delta\eta)^2+(\Delta\phi)^2}<0.4$ around its direction in
pseudorapidity and azimuth is defined. A photon is considered to be isolated if the sum of the \ET of all
the other final state particles produced from the same hard scattering inside the isolation cone is smaller than 5\GeV.
The \GEANTfour simulation is used to
determine the isolated photon energy and efficiency corrections.

\begin{figure}[htbp]
\begin {center}
\includegraphics[width=0.44\textwidth]{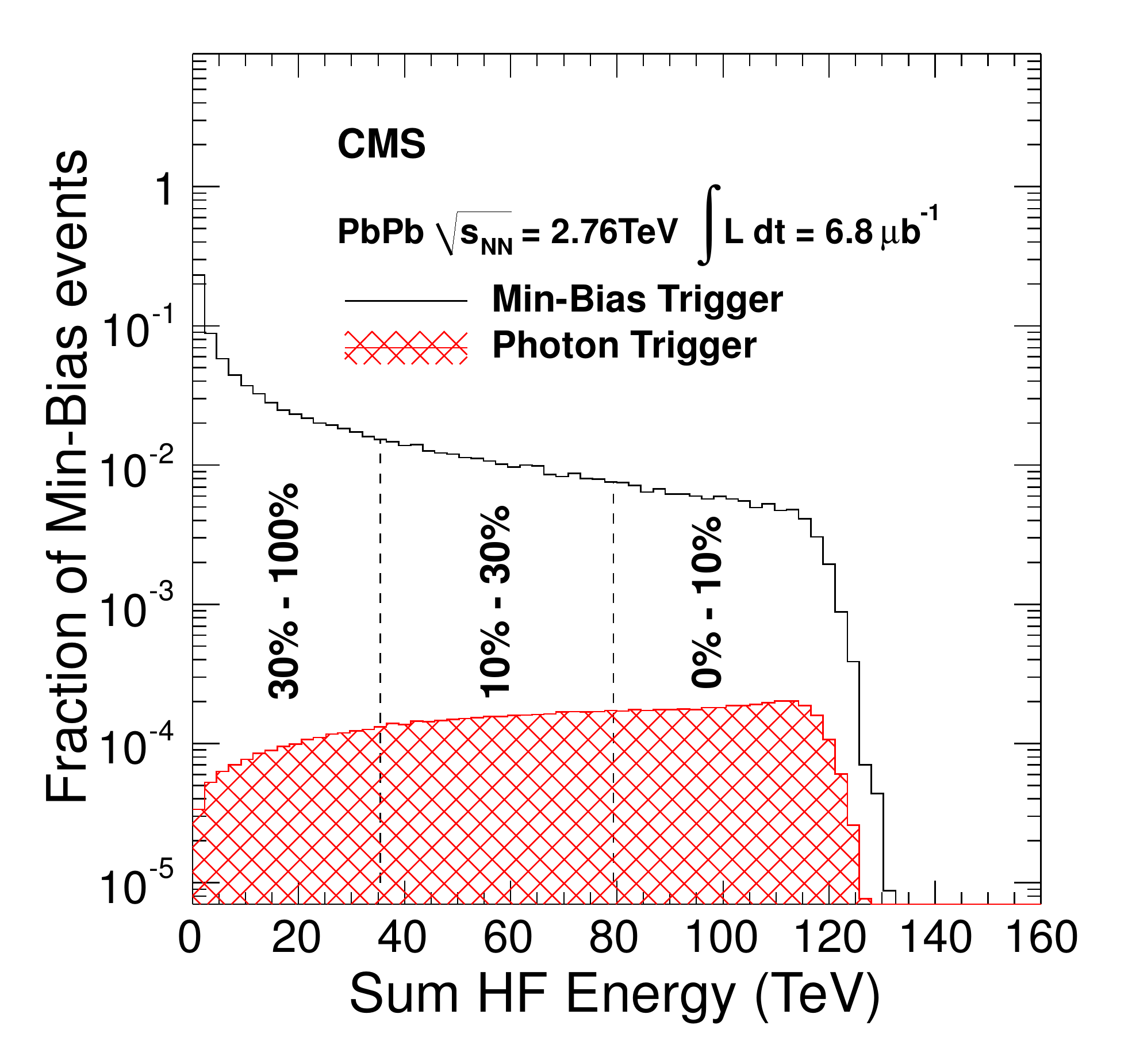}
\caption{
Probability distribution of the total HF energy for minimum-bias PbPb collisions (black open histogram). The three regions separated by the vertical dotted lines correspond to the centrality ranges used in this analysis. Also shown is the HF energy distribution for the subset of events passing the HLT photon trigger (cross-hatched histogram), which is about 3.3\% of all minimum-bias events.}
\label{fig:HF_cent}
\end{center}
\end{figure}

\section{PbPb centrality determination}
\label{section:centrality}

For the analysis of PbPb events, it is important to determine the overlap or impact parameter of the two colliding
nuclei, usually called the reaction ``centrality''. Centrality is determined with the minimum-bias
sample using the total sum of energy signals from the HF. 
The PbPb MB data sample is divided into three percentile ranges of the total inelastic cross section: 0--10\% (most central, small impact parameter), 10--30\% (mid-central), and 30--100\% (peripheral, large impact parameter). The distribution of the HF energy, along with the intervals defining the three event classes,
are shown in Fig.~\ref{fig:HF_cent}. Details of the centrality determination are described in
Ref.~\cite{Chatrchyan:2011sx}. The intervals can be correlated with geometrical properties of the collision
using a Glauber model simulation~\cite{Miller:2007ri}.
The two most commonly used quantities are N$_\text{part}$, the total number of nucleons in the two Pb nuclei that
experience at least one collision, and N$_\text{coll}$, the total
number of inelastic nucleon-nucleon collisions. The variable $N_\text{part}$ is often
used to quantify the reaction centrality, with $N_\text{part}$ = 2 corresponding
to a single nucleon-nucleon interaction and $N_\text{part}$ = $2 \times 208$ corresponding
to a head-on PbPb collision where all nucleons participate. The variable
$N_\text{coll}$ quantifies the total number of incoherent nucleon-nucleon
collisions at a given centrality, and since this is directly proportional
to the high-\pt particle production yields, $N_\text{coll}$ is used to normalize
the PbPb yields for comparison with the same observables for hard processes measured
in \Pp\Pp\ collisions. As can be seen in Fig.~\ref{fig:HF_cent}, the centrality distribution associated with hard processes, such as high-\et photon production (cross-hatched histogram), has a more pronounced contribution from central collisions than for minimum-bias events (solid line)

\section {Photon reconstruction and identification}
\label{photonRecoID}

\begin{figure*}[htbp]
\begin{center}
\includegraphics[width=0.95\textwidth]{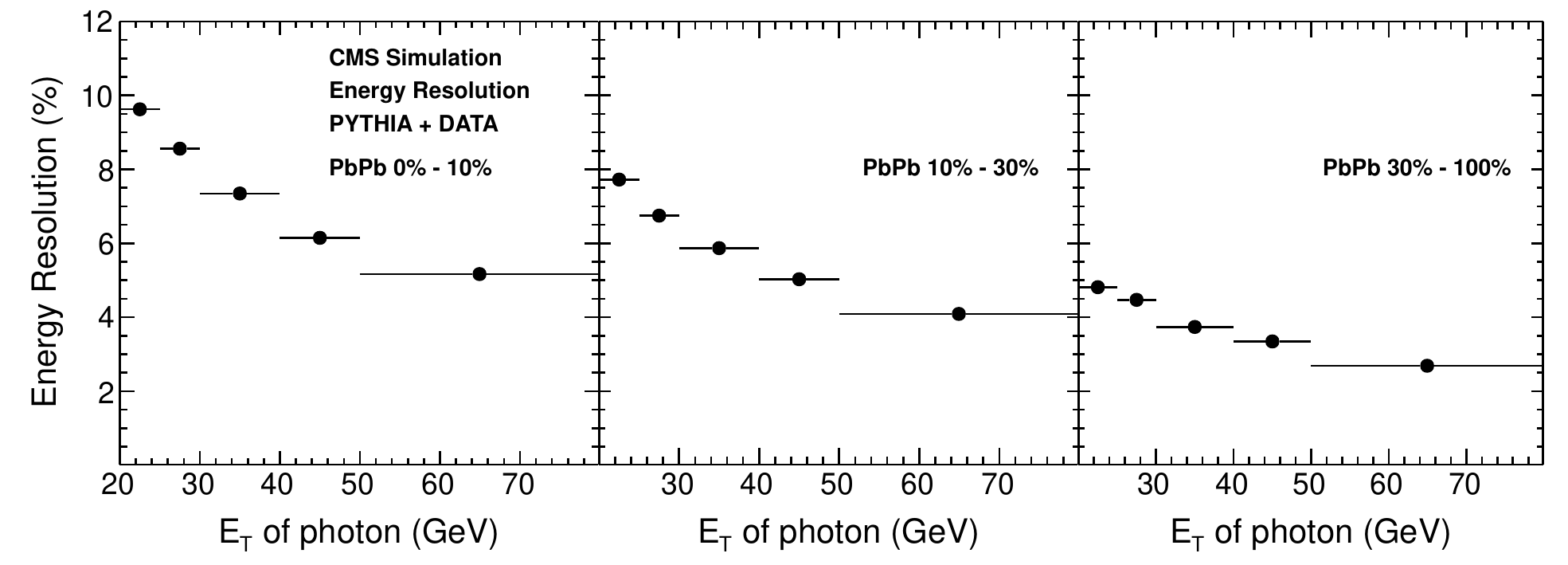}
\caption{\label{fig:energyResolution}Relative energy resolution of reconstructed photons as a function of photon transverse energy, determined using $\Pgg$+jet \textsc{pythia+data} sample for three centrality intervals.  The horizontal bars indicate the bin width.}
\end{center}
\end{figure*}

The photon reconstruction algorithm and isolation requirements in \Pp\Pp\ collisions are detailed in Ref.~\cite{PhysRevD.84.052011}. The reconstruction
in PbPb collisions is very similar, although some modifications are introduced in order to deal with
the large background of particles produced in the collision.
ECAL ``superclusters'' are reconstructed in the barrel region of the electromagnetic calorimeter using the
``island'' energy-clustering algorithm~\cite{PTDR2}. The first step of the algorithm is a
search around the seeds, which are defined as cells (reconstructed hits) with a transverse energy above a
threshold of 0.5\GeV. Starting from a seed position, adjacent cells are examined, scanning
first in the $\phi$ and then in the $\eta$ direction.  Cells are added to the cluster until the cell under consideration satisfies one of three conditions; the corrected energy deposit in the cell is zero, the energy in the cell is larger than in the adjacent cell which was already added to the cluster, or the cell is already part of a different island cluster. In the second step, the island clusters are merged into superclusters. The procedure is seeded
by searching for the most energetic cluster above a transverse energy threshold ($\ET > 1\GeV$) and then
collecting all the other nearby clusters that have not yet been used in a narrow $\eta$-window
($\Delta\eta$ = 0.07), and a much wider $\phi$-window ($\Delta\phi$ = 0.8).
A photon candidate is constructed from a ``supercluster'' (conglomerate of energy deposits)
with uncalibrated $\ET> 8\GeV$, and its energy is corrected to account for the material in front
of the ECAL and for electromagnetic shower containment. The direction of the photon is also recalculated with respect to the primary vertex.
An additional energy correction is applied to remove the background contribution from the underlying PbPb
event. This correction is obtained from the $\Pgg$+jet \textsc{pythia+data} sample and listed in
Table~\ref{table:energyScaleTable} for the 3 centrality intervals. The underlying PbPb activity also worsens the photon energy resolution
to a maximum of 9\% for the lowest \ET bin in the 0--10\% central events, as shown in Fig.~\ref{fig:energyResolution}.

\begin{table}[htbp]
  \centering
  \caption{\label{table:energyScaleTable}Energy correction factors for the background energy contribution found using the
$\Pgg$+jet \textsc{pythia+data} sample for each centrality interval and photon \ET.
The reconstructed \ET of photon candidates with $|\eta^\gamma|<1.44$ is multiplied by
this factor to get the corrected transverse energy $E^\gamma_\text{T}$.}
  \begin{tabular}{|c|ccc|}
    \hline
    \hline
    Photon \ET  & & PbPb centrality & \\
    (GeV) & 0--10\% & 10--30\% & 30--100\% \\
    \hline
    20--25 & 0.90 & 0.94 & 0.99  \\
    25--30 & 0.91 & 0.95 & 0.99  \\
    30--40 & 0.92 & 0.95 & 0.99 \\
    40--50 & 0.94 & 0.96 & 0.99 \\
    50--80 & 0.95 & 0.97 & 0.99  \\
    \hline
    \hline
  \end{tabular}
\end{table}

Anomalous signals caused by the interaction of heavily ionizing particles directly with the silicon avalanche photodiodes used for the ECAL barrel readout are removed by the following requirements: (i) the signal should be consistent in time (within 3\unit{ns}) with a
photon from the collision; (ii) the sum of the energy in the four adjacent cells surrounding the central
cell should be at least 10\% of the central cell energy. These two selections are satisfied by 99.7\%
of the photon signal candidates.

The selected photon candidates are required to be in the ECAL barrel within the pseudorapidity interval
$\abs{\eta^\Pgg}<1.44$, to not match with any electron candidates in a search window of $\abs{\eta^\Pgg-\eta^\text{Track}}<0.02$ and $\abs{\phi^\Pgg-\phi^\text{Track}}<0.15$ with respect to the associated electron candidate
track, and to have $E^\gamma_\mathrm{T} > 20\GeV$.  A first rejection of neutral mesons mimicking a high-\ET photon
candidate in the ECAL is done using the $H/E$ ratio defined as the ratio of hadronic energy to
electromagnetic energy inside a cone of $\Delta R = 0.15$, computed from the energy depositions in the HCAL and the ECAL~\cite{CMSppPhoton}.
Photon candidates with $H/E < 0.2$ are selected for this analysis.

To measure the isolation of a given photon candidate in a PbPb event, the detector activity in a cone of radius $\Delta R=0.4$ with respect to the centroid of
the cluster is used. Calorimeter-based isolation variables $\mathrm{Iso_{ECAL}}$ and $\mathrm{Iso_{HCAL}}$
are calculated by summing over the ECAL and HCAL transverse energy, respectively, measured inside the cone, while a
track-based isolation variable $\mathrm{Iso_{Track}}$ is measured by summing over the transverse momentum of
all tracks with $p_\text{T}>2\GeVc$ inside the cone. The total ECAL energy associated with the photon candidate is excluded in the
$\mathrm{Iso_{ECAL}}$ calculation. In order to remove the contribution of hadronic activity from the  underlying
PbPb event background falling inside the isolation cone for each centrality, the average value of the
energy deposited per unit area in the $\eta-\phi$ phase space ($\langle \mathrm{UE} \rangle$) is estimated within
a rectangular region $2\Delta R$-wide and centered on $\eta^\Pgg$ in the $\eta$-direction and  2$\pi$ wide in the
$\phi$-direction, excluding the isolation cone. The UE-subtracted isolation variables
$\mathrm{Iso^{UE-sub}} = \mathrm{Iso} - \pi(\Delta R)^2 \langle \mathrm{UE} \rangle$ are used to further reject photon
candidates originating from jets. The sum of the isolation variables
$(\mathrm{SumIso^{UE-sub} = Iso^{UE-sub}_{ECAL} + Iso^{UE-sub}_{HCAL} + Iso^{UE-sub}_{Track}})$ is required to be smaller
than 5\GeV. The efficiency of the isolated photon identification criteria in PbPb collisions, which is obtained from the \textsc{pythia+data} sample, is summarized in Table~\ref{table:efficiencyTable}.

\begin{table*}[htbp]
  \centering
  \caption{\label{table:efficiencyTable}  Efficiencies of the isolated photon identification at each step: clustering,
    anomalous signal removal, $H/E$ selection, and isolation requirement.  Numbers in each row are the
    efficiencies relative to the previous step.  The selections are more efficient for high $E^\gamma_\mathrm{T}$
    photons and for more peripheral events. The intervals given
    indicate the $E^\gamma_\mathrm{T}$-dependent variations of the efficiencies.}
\begin{tabular}{|c|ccc|}
    \hline
    \hline
    & & PbPb centrality & \\
    Isolated photon identification & 0--10\% & 10--30\% & 30--100\% \\
    \hline
    Supercluster reconstruction & 96--99\%  & 97--99\%  & 97--99\% \\
    Anomalous signal removal   & 99--100\% & 99--100\% & 99--100\% \\
    $H/E$ $<$ 0.2                 & 96--99\% & 98--99\% & 99--100\%  \\
    $\mathrm{SumIso^{UE-sub}}<5\GeV$ & 82--84\%& 86--88\% & 96--97\% \\
    \hline
    Total                 & 77--82\%  & 83--86\% & 92--95\% \\
    \hline
    \hline
  \end{tabular}

\end{table*}

\section{Signal extraction, corrections, and systematic uncertainties}
\label{sec:signal_ext}

The selection criteria described above yield a relatively pure sample of isolated photons.
However, there are still
non-prompt photons, such as those from isolated $\Pgpz$s that are carrying a large fraction of the parent
fragmenting parton energy, which can pass the isolation cuts.
Those remaining backgrounds are estimated using a two-component fit of the shape of the electromagnetic
shower in the ECAL
and separated from the signal on a statistical basis, as described below.

The topology of the energy deposits can be used as a powerful tool to distinguish the signal from the
background by making use of the fine $\eta$ segmentation of the electromagnetic calorimeter.
The shower shape is characterized by a transverse shape variable $\sigma_{\eta \eta}$, defined as a modified second moment of the
electromagnetic energy cluster distribution around its mean $\eta$ position:
\begin{eqnarray}
\label{sieieFormula}
\sigma_{\eta \eta}^2 = \frac{\sum_i w_i(\eta_i-\bar{\eta})^2}{\sum_i w_i},~~~~ w_i = \mathrm{max}(0, 4.7 + \ln \frac{E_i}{E}),
\end{eqnarray}
where $E_i$ and $\eta_i$ are the energy and position of the $i^{th}$ crystal in a group of $5\times 5$ crystals
centered on the one with the highest energy, $E$ is the total energy of the crystals in the calculation and $\bar{\eta}$ is the average
$\eta$ weighted by $w_i$ in the same group~\cite{positionLog,CMSppPhoton}. Isolated photons tend to have a
smaller mean value of $\sigma_{\eta \eta}$ and a narrow distribution, while photons produced in hadron decays
tend to have larger $\sigma_{\eta  \eta}$ mean and a wider $\sigma_{\eta  \eta}$ distribution.

The isolated prompt photon yield is estimated with a binned maximum likelihood fit to the $\sigma_{\eta \eta}$
distribution with the expected signal and background components for each $E^\gamma_\text{T}$ interval.
The signal and background component shapes used in the \Pp\Pp\ analysis are described in~\cite{CMSppPhoton}.
In the PbPb analysis, the signal component shape for each $E^\gamma_\text{T}$ and centrality bin is obtained from
$\Pgg$+jet \textsc{pythia+data} samples, and the background component shape is extracted from data using
a background-enriched SumIso sideband ($6<\mathrm{SumIso^{UE-sub}}<11\GeV$) sample while keeping all other selection criteria unchanged.

\begin{figure*}[htbp]
  \begin{center}
    \includegraphics[width=0.95\textwidth]{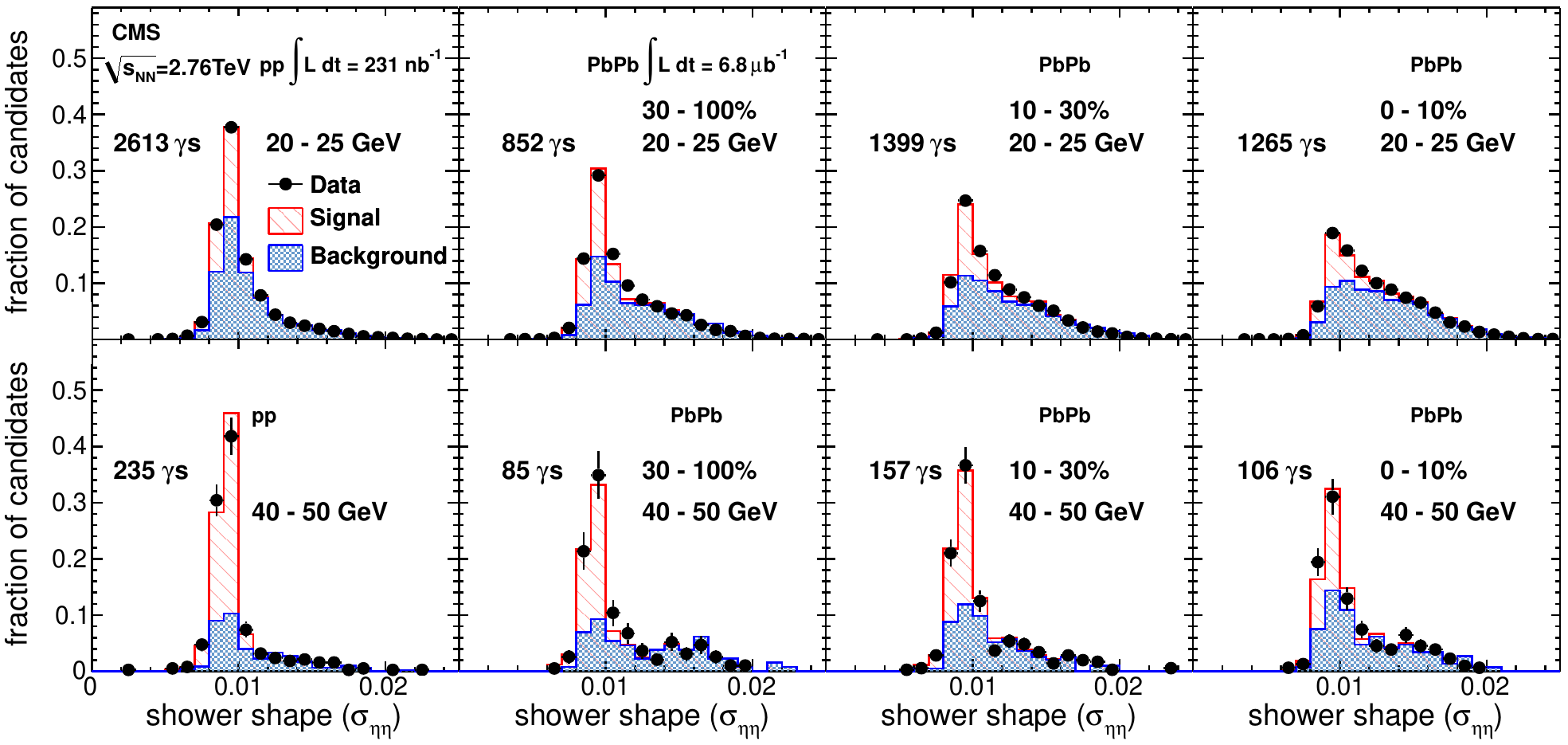}
    \caption{\label{fig:fit} Measured shower-shape $\sigma_{\eta \eta}$ distribution for photon candidates with
   $\ET^\gamma = \mbox{20--25}\GeV$ and 40--50\GeV in \Pp\Pp\ (2 left plots) and PbPb collisions for 3 different centrality ranges. The extracted numbers of isolated photons are shown in the figure. The fit result (red line), signal (red-hatched histogram) and background components (blue shaded histogram) are also shown. }
  \end{center}
\end{figure*}

Figure ~\ref{fig:fit} illustrates the results of the two-component fit of the shower-shape distribution
measured in \Pp\Pp\ and PbPb collisions. The remaining background contribution from electrons passing all the photon
selection criteria, estimated from a sample enriched in isolated electrons found by reversing the electron-veto requirement described in Section~\ref{photonRecoID}, is also subtracted
to extract the raw signal yields $(N^{\gamma}_\text{raw})$. Typically, the contribution due to electron contamination in PbPb collisions is 3--6\% for different \et intervals.
A bin-by-bin correction for the energy smearing ($U$), which amounts to 1.00--1.08 for different $E^\gamma_\text{T}$
and centrality bins, is also applied to the raw signal yields to obtain the number of isolated photons. The \et-differential photon yield per event is defined as
\begin{eqnarray}
\frac{dN^{\gamma}_\mathrm{PbPb}}{dE^\gamma_\text{T}} = \frac{N^{\gamma}_\text{raw}}{U\times\epsilon \times f_\text{cent}\times N_\text{MB} \times \Delta E^\gamma_\text{T}},
\end{eqnarray}
where $N_\text{MB}$ is the number of sampled minimum-bias PbPb events, $f_\text{cent}$ is the fraction of PbPb events in each centrality bin, and $\epsilon$ is the efficiency
of the isolated photon identification (Table~\ref{table:efficiencyTable}). For \Pp\Pp\ collisions we
normalize the yields by the integrated luminosity (${\cal L}_{\Pp\Pp}$) to obtain the \et-differential cross section $d\sigma^{\gamma}_{\Pp\Pp}/dE^\gamma_\text{T}$:
\begin{eqnarray}
\frac{d\sigma^{\gamma}_{\Pp\Pp}}{dE^\gamma_\text{T}}=\frac{N^{\gamma}_\text{raw}}{U\times\epsilon\times {\cal L}_{\Pp\Pp} \times \Delta E^\gamma_\text{T}}.
\end{eqnarray}

\begin{table*}[htbp]
 \centering
 \caption{\label{table:SysSieieSignal}  Summary of the contributions to the estimated systematic
   uncertainties on the isolated photon spectra measured in \Pp\Pp\ and PbPb collisions and their total. The nuclear overlap function $T_\text{AA}$ is defined in Section~\ref{sec:results}.
 The intervals indicate the $E^\gamma_\text{T}$-dependent variations of the uncertainties.}
 \begin{tabular}{|c|c|ccc|}
   \hline
   \hline

   & \Pp\Pp  & & PbPb centrality &   \\
   Source & & 0--10\% & 10--30\% & 30--100\%  \\
   \hline
    Efficiency &  1--5\% & 5--9\% & 5--7\% & 5--6\%  \\
    Signal modeling & 3--5\% & 1--5\%  & 3--5\% & 1--4\%  \\
    Background modeling & 9--13\% & 15--23\% & 14--16\% & 12--21\% \\
    Electron veto & 1\% &  3--6\%  & 3--5\% & 3--5\% \\
    Photon isolation definition & 2\% & 7\% & 5\% & 2\%  \\
    Energy scale  & 3--6\% & 9\% & 9\% & 9\%  \\
    Energy smearing & 1\% & 4\% & 4\% & 4\% \\
    Shower-shape fit & 3\% & 5\% & 5\% & 5\% \\
    Anomalous signal cleaning & 1\% & 1\% & 1\% & 1\% \\
    $N_\text{MB}$ & -- & 3\% & 3\% & 3\% \\
    Luminosity & 6\% \ & -- & -- & -- \\
    \hline
    Total without $T_\text{AA}$ & 14--16\% & 23--30\% &  22--25\% &  23--28\% \\
   \hline
    $T_\text{AA}$ & -- & 4\% &  6\% &  12\% \\
   \hline
    Total  & 14--16\%  & 23--30\% &  23--26\% &  26--31\% \\
   \hline
   \hline
 \end{tabular}
\end{table*}

The systematic uncertainties of the measured photon spectra are summarized in Table~\ref{table:SysSieieSignal}. The total systematic
uncertainties are 22--30\% for PbPb and 14--16\% for \Pp\Pp\ collisions.
The systematic uncertainty of the photon yield $dN^{\gamma}_\mathrm{PbPb}/dE^\gamma_\text{T}$ in PbPb collisions is dominated by the uncertainty on the background modeling. Since the transverse shape variable $\sigma_{ \eta \eta}$ may be correlated with the number of particles in the isolation cone (characterized by ${\mathrm SumIso^{UE-sub}}$), non-prompt photons from \textsc{pythia+data} samples are used to examine the possible difference between the $\sigma_{ \eta \eta}$ distribution in the $\mathrm{SumIso^{UE-sub}}$ signal and in the sideband regions. Systematic checks are performed using the differences in the mean and width seen in the MC to vary the background component shape in the fit. The estimated uncertainty is in the range of 12--23\%, where the given interval indicates the $E^\gamma_\text{T}$ and centrality-dependent variations of the uncertainty.
The uncertainty due to the $\sigma_{ \eta \eta}$ distribution of isolated photons is estimated by comparing the distributions of electrons from MC and data.
Given the small number of $\cPZ\rightarrow \Pep\Pem$ events in the PbPb data sample, $\cPZ\rightarrow \Pep\Pem$ events from the 2010 \Pp\Pp\ run at $\sqrt{s} = 7\TeV$ are mixed with MB PbPb
data. The differences in the measured mean and width from those obtained in the MC($\cPZ\rightarrow \Pep\Pem$)+PbPb
data are used to vary the $\sigma_{ \eta \eta}$ distributions of isolated photons. Such systematic changes
result in a final propagated uncertainty of 1--5\% in the isolated photon yield.
The uncertainty due to the energy scale propagates to an uncertainty of 9\% in the final spectra.
The uncertainty due to the energy smearing correction is
obtained by varying the assumed isolated photon differential cross section at low photon \et (used to obtain the unfolding
correction factors) by $\pm$50\%, and is found to be 4\%. The uncertainty of the two-component fit is checked by
using different binning widths in the fit, and is found to be 5\%. A 3--6\% uncertainty is associated
with the electron contamination subtraction.
The difference between experimental and theoretical photon isolation definitions as described in Section~\ref{sec:Theory} due to the detector response and underlying event is estimated to be 2--7\%. The uncertainty of $N_\text{MB}$ due to the MB selection efficiency is 3\% in PbPb collisions, and
a 6\% uncertainty is quoted for the integrated luminosity in \Pp\Pp\ collisions.

\section{Theoretical calculations}
\label{sec:Theory}

The isolated photon spectra measured in \Pp\Pp\ and PbPb collisions at $\snn = 2.76\TeV$
are compared with next-to-leading-order (NLO) pQCD predictions obtained using \textsc{jetphox} 1.2.2,
which reproduces well the measured \Pp\Pp\ data at $\sqrt{s} = 7\TeV$~\cite{CMSppPhoton,PhysRevD.84.052011}.
The inclusive photon spectrum in \Pp\Pp\ collisions is computed using the CT10~\cite{CT10} parton distribution functions. The same spectrum in PbPb collisions is computed using the NLO EPS09~\cite{Eskola:2009uj} PDFs, which include nuclear modifications of the proton PDFs. The reduction of photon emission due to isospin
effects in nuclear compared to proton collisions (the relative population of u and d quarks is not identical in a single proton and in a lead nucleus, with 126 neutrons and 82 protons) is accounted
for in the calculations.
For both systems the BFG-II set~\cite{BFGII} of parton-to-photon fragmentation functions is used, and
the default renormalization, factorization, and fragmentation scales ($\mu_{R}$, $\mu_{F},$ and $\mu_{f}$) are all set
to the photon \ET. The parton-level isolation, summing over the transverse energy of all partons inside a
cone of radius $\Delta R = 0.4$, is required to be smaller than 5\GeV.
In order to estimate the dependence of the predictions on the choice of theoretical scales, the
$\mu_{i}$ scales are varied by a factor of 2 below and above their default values, keeping the ratio between any two
scales less than or equal to 2.
The uncertainty linked to the choice of the proton PDF is $\pm$(7--5)\%
and it is smaller than the theoretical scale uncertainty which varies
within $\pm$(15--10)\% in the measured \pt range, as found in~\cite{CMSppPhoton}.
The uncertainty on the predictions due to nuclear PDFs is estimated using
the 30 eigenvalues of the EPS09 PDF set. In addition, the PbPb spectrum is computed using two alternative nuclear
PDF sets: nDS~\cite{deFlorian:2003qf} and HKN07~\cite{Hirai:2007sx}. When data are compared to \Pp\Pp\ NLO
predictions, the proton PDF and the scale uncertainties are added in quadrature.

\section{Results}
\label{sec:results}

In order to compare the cross sections for any high-\pt particle produced in PbPb and \Pp\Pp\ collisions, a scaling factor, the nuclear overlap function $T_\text{AA}$, is needed to provide proper normalization. This factor, equal to the number of nucleon-nucleon (NN) collisions, $N_\text{coll}$, normalized by the pp inelastic cross section, can be interpreted as the NN-equivalent integrated luminosity at any given PbPb centrality. The LHC collaborations use a common nucleon-nucleon inelastic cross section of $\sigma$~=~64~$\pm$~5\unit{mb} at 2.76\TeV, based on a fit of the existing data for total and elastic cross sections in proton-proton and proton-antiproton collisions~\cite{pdg}. In units of $\mathrm{mb}^{-1}$, the average values of $T_\text{AA}$ are $23.2\pm 1.0$, $11.6 \pm 0.7$, $1.45\pm 0.18$, and
$5.66\pm 0.35$ for the centrality ranges 0--10\%, 10--30\%, 30--100\%, and 0--100\%, respectively. These numbers are
computed with a Glauber model~\cite{Miller:2007ri} using the same parameters as in~\cite{Chatrchyan:2011sx}.
The quoted uncertainties are derived by varying the Glauber model parameters and the MB trigger and event selection efficiency within their uncertainties. The measured $\ET^\gamma$-differential isolated photon cross sections in \Pp\Pp\ and the $T_\text{AA}$-scaled yields in PbPb
collisions, including both statistical and systematic uncertainties, are listed in Table~\ref{table:crossSectionTable}.

\begin{table*}[htbp]
 \centering
 \caption{\label{table:crossSectionTable}
Isolated photon cross sections for $|\eta^\gamma|<1.44$ in bins of $\ET^\gamma$ for \Pp\Pp\ collisions and PbPb
collisions (for 3 centrality intervals and for the full range) at $\snn=2.76\TeV$.
The first uncertainty is statistical and the second one is systematic (including
$T_\text{AA}$ uncertainties in the PbPb case).}
 \begin{tabular}{|c|c|ccc|c|}
   \hline
   \hline
  \ET & \multirow {2}{*}{\footnotesize \Pp\Pp\ $d\sigma^{\gamma}_{\Pp\Pp}/dE^\gamma_\text{T}$ (pb/GeV)} & \multicolumn{4}{c|}{\footnotesize PbPb    $dN^{\gamma}_\mathrm{ PbPb}/dE^\gamma_\text{T}/\langle T_\text{AA}\rangle$ (pb/GeV)} \\
   \cline{3-6}
(GeV) & & 0--10\% & 10--30\% & 30--100\% & 0--100\%  \\
   \hline
  20--25 & {\footnotesize2400 $\pm$140 $\pm$400} & {\footnotesize2480 $\pm$240 $\pm$740} & {\footnotesize2560 $\pm$210 $\pm$620} & {\footnotesize3310 $\pm$280 $\pm$950} & {\footnotesize2660 $\pm$140 $\pm$810} \\
  25--30 & {\footnotesize983 $\pm$74 $\pm$159} & {\footnotesize830 $\pm$120 $\pm$240} & {\footnotesize1110 $\pm$120 $\pm$250} & {\footnotesize1220 $\pm$170 $\pm$350} & {\footnotesize1013 $\pm$75 $\pm$292} \\
   30--40  & {\footnotesize305 $\pm$30 $\pm$45} & {\footnotesize416 $\pm$54 $\pm$110} & {\footnotesize383 $\pm$46 $\pm$85} & {\footnotesize353 $\pm$60 $\pm$94} & {\footnotesize391 $\pm$31 $\pm$105} \\
   40--50  & {\footnotesize102 $\pm$12 $\pm$15} & {\footnotesize100 $\pm$22 $\pm$23} & {\footnotesize142 $\pm$21 $\pm$32} & {\footnotesize161 $\pm$30 $\pm$43} & {\footnotesize128 $\pm$14 $\pm$33} \\
   50--80 & {\footnotesize20.1 $\pm$2.6 $\pm$2.8}  & {\footnotesize20.0 $\pm$5.7 $\pm$4.6} & {\footnotesize21.8 $\pm$5.5 $\pm$5.0} & {\footnotesize24.3 $\pm$4.9 $\pm$5.6} & {\footnotesize21.5 $\pm$3.4 $\pm$5.0} \\
   \hline
    \hline
 \end{tabular}
\end{table*}

\begin{figure}[htbp]
\begin{center}
\includegraphics[width=0.48\textwidth]{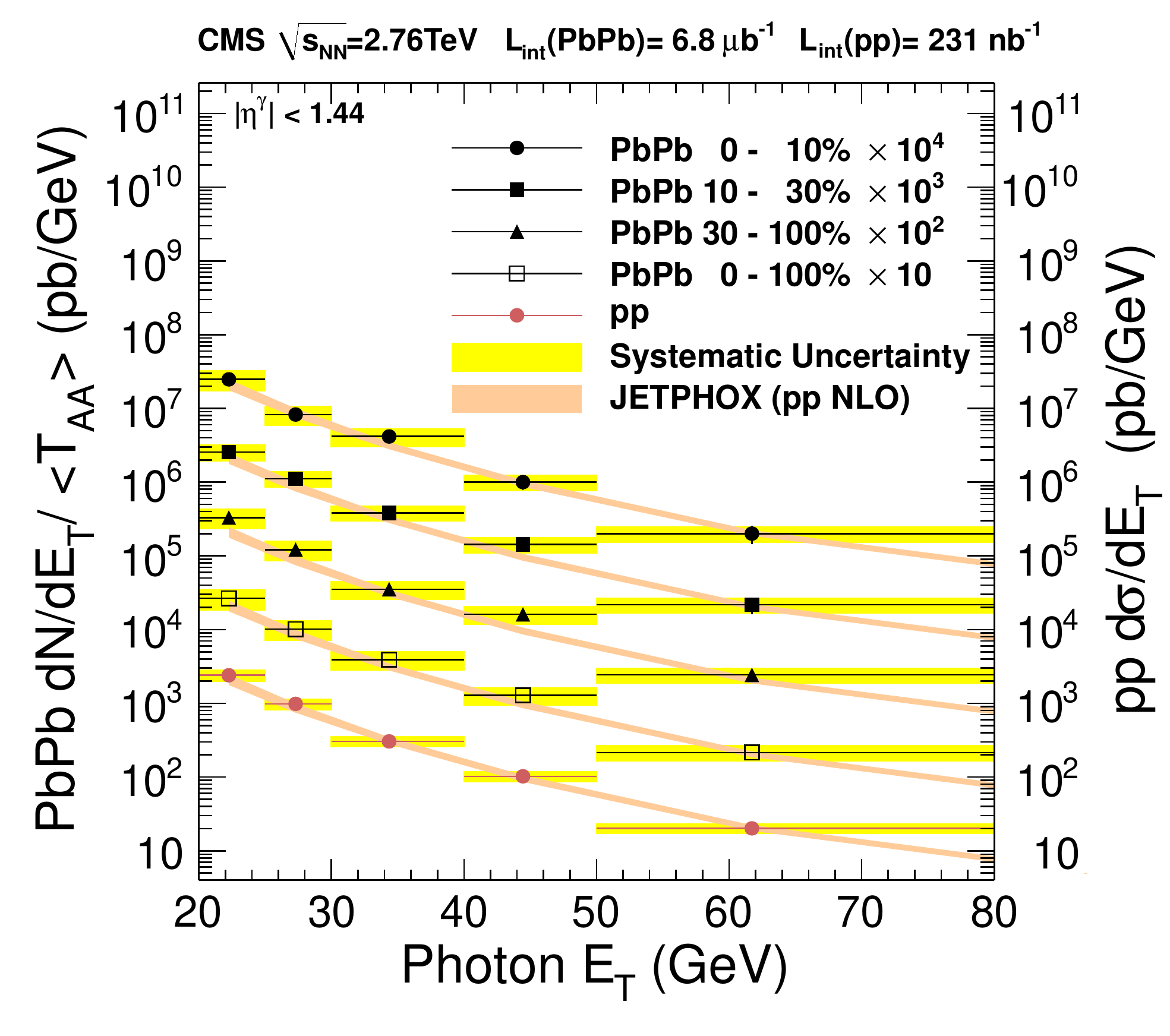}
\caption{\label{fig:dndet} Isolated photon spectra measured as a function of $E^\gamma_\text{T}$ for 0--10\%, 10--30\%, 30--100\%,
  0--100\% PbPb collisions (scaled by $T_\text{AA}$) and \Pp\Pp\ collisions at 2.76\TeV, scaled by
  the factors shown in the figure for easier viewing. The horizontal bars indicate the bin width. The total systematic uncertainty (bottom row of Table~\ref{table:SysSieieSignal})
  is shown as a yellow box at each \et bin. The results are compared to the NLO \textsc{jetphox} calculation (see
  text) with its associated scale and PDF uncertainties (added in quadrature) shown as a pink band. }
\end{center}
\end{figure}

Figure~\ref{fig:dndet} shows the \Pp\Pp\ cross sections
and the PbPb
$T_\text{AA}$-scaled yields
compared to the \textsc{jetphox} predictions obtained with
the CT10 PDF, described in Section~\ref{sec:Theory}. The data are plotted at the true centre of the $E^\gamma_\text{T}$ distributions in each bin~\cite{trueCenter}.
The \Pp\Pp\ and PbPb data are consistent with the NLO calculation
at all transverse energies within the quoted statistical and systematic uncertainties.

\begin{figure}[htbp]
\begin{center}
\includegraphics[width=0.48\textwidth]{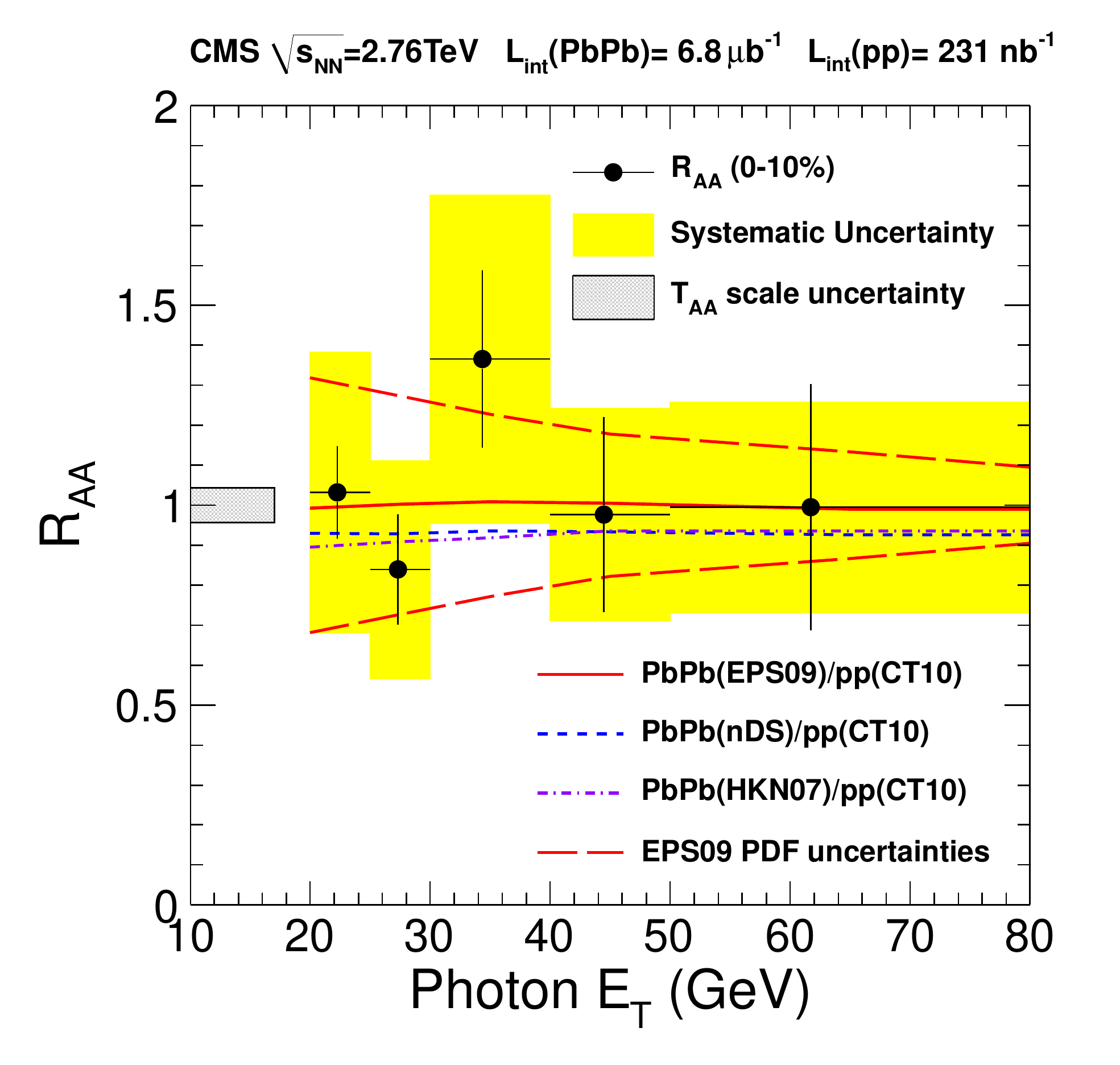}
\caption{\label{fig:RAA} Nuclear modification factor $R_\text{AA}$ as a function of the photon \et measured in
  the 0--10\% most central PbPb collisions at 2.76\TeV.  The vertical error bars indicate the statistical uncertainty and the horizontal bars reflect the bin width. The total systematic uncertainties without the $T_\text{AA}$ uncertainty (see Table~\ref{table:SysSieieSignal}) are shown as yellow
  filled boxes. The $T_\text{AA}$ uncertainty, common to all points, is indicated by the left box centered at unity.
  The curves show the theoretical predictions, obtained with
  \textsc{jetphox} for various nuclear PDFs described in the text. The uncertainty from the EPS09 PDF parameters is shown as the red dashed lines.}
\end{center}
\end{figure}

The nuclear modification factor ($R_\text{AA}$) for isolated photon production in PbPb collisions,
\begin{eqnarray}
R_\text{AA}= dN^{\gamma}_\text{PbPb}/dE^\gamma_\text{T}/(T_\text{AA}\times d\sigma^{\Pgg}_{\Pp\Pp}/d\ET),
\end{eqnarray}
is computed from the measured PbPb scaled yield
for each centrality and the \Pp\Pp\ differential cross section.
Figure~\ref{fig:RAA} displays $R_\text{AA}$ as a function of the isolated photon \ET for the 0--10\% most central
PbPb collisions. The ratio is compatible with unity within the experimental uncertainties for all
\et values. This confirms the validity of the $T_\text{AA}$ scaling expectation for perturbative cross sections
in nucleus-nucleus collisions at the LHC, as found previously for $Z$-boson production~\cite{Chatrchyan:2011ua}.
Changes in the isolated photon yields in PbPb collisions compared to \Pp\Pp\ due to modifications of the nuclear parton
densities are relatively small in this high-\et range, according to the \textsc{jetphox} calculations.
Figure~\ref{fig:RAA} shows that the calculated NLO ratios of the PbPb to \Pp\Pp\ isolated photon spectra obtained with the central
values of the EPS09, nDS and HKN07 nuclear PDFs differ at most by $\pm$10\%. The band of uncertainty obtained
from the 68\% confidence level variation of the EPS09 nuclear parton distribution parameters (red dashed lines) is
fully consistent with the measured nuclear modification factor at all transverse energies.

\begin{figure*}[htbp]
\begin{center}
\includegraphics[width=0.95\textwidth]{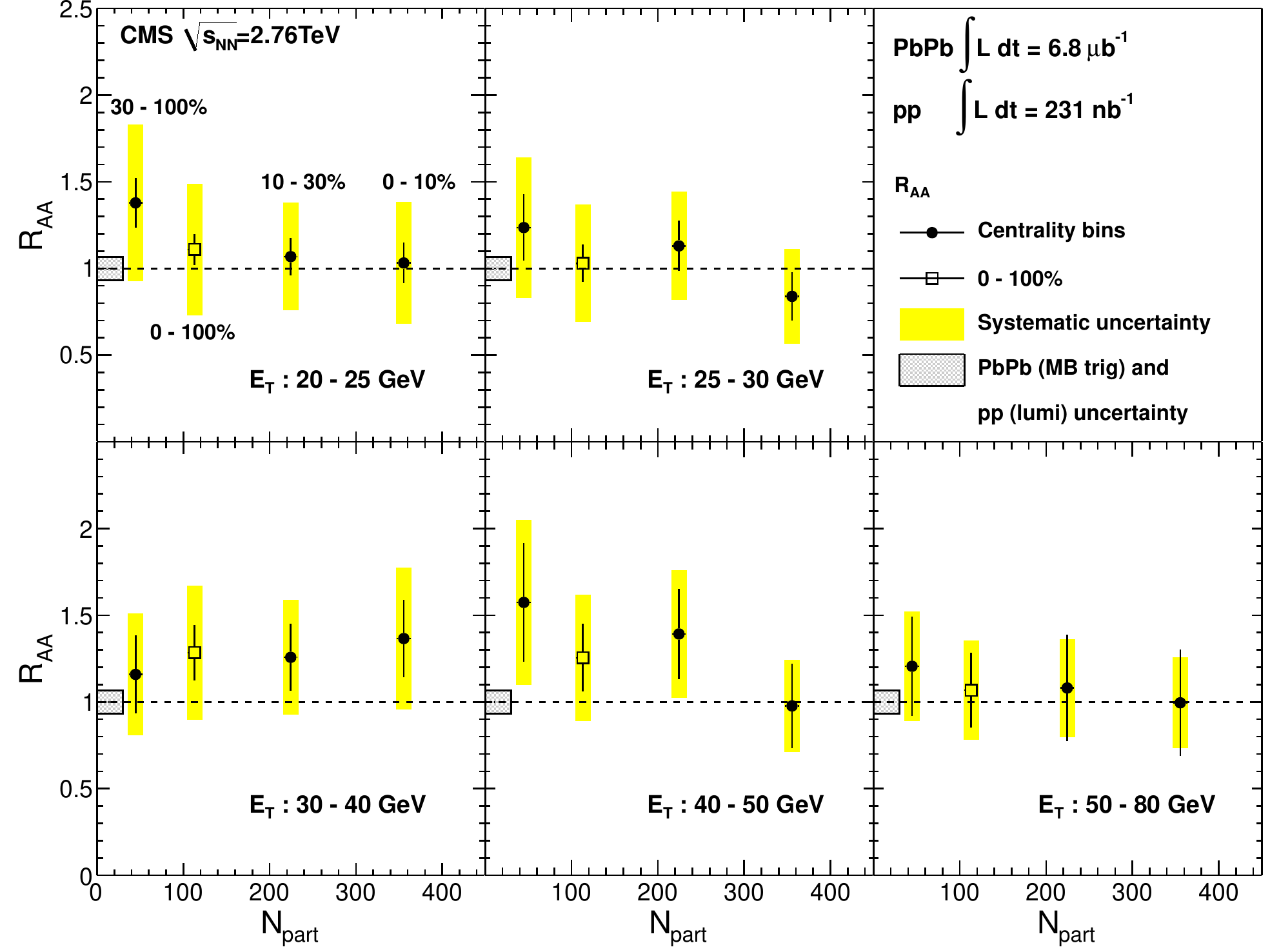}

\caption{\label{fig:RAAvsNpart} The measured nuclear modification factor $R_\text{AA} $ as a function of PbPb centrality
  (given by the number of participating nucleons, $N_\text{part}$) for five different photon transverse
  energy intervals. The error bars on each point indicate the statistical uncertainty. The systematic uncertainties are shown
  as yellow boxes, including the centrality dependent $T_\text{AA}$ uncertainty. The common uncertainties related to event selection efficiency and \Pp\Pp\
 integrated luminosity are shown as grey hatched boxes around unity.}
\end{center}
\end{figure*}

In order to investigate the centrality dependence of the isolated photon production yields in PbPb compared to
\Pp\Pp\ collisions, Fig.~\ref{fig:RAAvsNpart} plots the $R_\text{AA}$ as a function of $N_\text{part}$ for various \ET bins.
Within the uncertainties, the measured nuclear modification ratio is consistent with unity, not only for minimum-bias PbPb collisions, but also for central collisions and all photon transverse energies. With improved statistical accuracy and/or reduced systematic uncertainties, isolated photon production yields in PbPb collisions at the LHC could be used to better constrain the nuclear PDFs by including the measurement in standard global fits of parton densities~\cite{Eskola:2009uj,deFlorian:2003qf,Hirai:2007sx}, as discussed in~\cite{Arleo:2011gc}.

\section{Summary}

In summary, the isolated photon spectra at midrapidity ($\abs{\eta^\Pgg}<1.44$)
have been measured as a function of transverse energy in \Pp\Pp\ and PbPb collisions at nucleon-nucleon centre-of-mass energies of 2.76\TeV.
The measured spectra are well reproduced by NLO
perturbative QCD calculations with recent parton distribution functions for the proton and nucleus.
No modification is observed in the $\ET^\Pgg$ spectra measured in PbPb collisions at various centralities
with respect to the \Pp\Pp\ differential cross sections scaled by the corresponding nuclear overlap function.
The result confirms the $T_\text{AA}$ scaling of perturbative cross sections in PbPb compared to \Pp\Pp\
collisions. It is consistent with the expectation that nuclear parton densities are not
significantly modified compared to the proton PDF in the explored kinematic range, dominated by high-\pt photons produced in parton-parton scatterings in the large-$Q^2$ and moderate parton fractional momentum $x$ region of the nuclear PDFs~\cite{Eskola:2009uj}.
Isolated photons are found to be unaffected by the produced strongly interacting medium, in sharp contrast to the large quenching effects observed for jets~\cite{Chatrchyan:2011sx}.
The measurement presented here establishes isolated photon production as a valuable perturbative probe of the
initial state in heavy-ion collisions and provides a baseline for the study of in-medium parton energy
loss in $\Pgg$+jet events.

\section*{Acknowledgements}

\hyphenation{Bundes-ministerium Forschungs-gemeinschaft Forschungs-zentren} We wish to congratulate our colleagues in the CERN accelerator departments for the excellent performance of the LHC machine. We thank the technical and administrative staff at CERN and other CMS institutes. This work was supported by the Austrian Federal Ministry of Science and Research; the Belgium Fonds de la Recherche Scientifique, and Fonds voor Wetenschappelijk Onderzoek; the Brazilian Funding Agencies (CNPq, CAPES, FAPERJ, and FAPESP); the Bulgarian Ministry of Education and Science; CERN; the Chinese Academy of Sciences, Ministry of Science and Technology, and National Natural Science Foundation of China; the Colombian Funding Agency (COLCIENCIAS); the Croatian Ministry of Science, Education and Sport; the Research Promotion Foundation, Cyprus; the Estonian Academy of Sciences and NICPB; the Academy of Finland, Finnish Ministry of Education and Culture, and Helsinki Institute of Physics; the Institut National de Physique Nucl\'eaire et de Physique des Particules~/~CNRS, and Commissariat \`a l'\'Energie Atomique et aux \'Energies Alternatives~/~CEA, France; the Bundesministerium f\"ur Bildung und Forschung, Deutsche Forschungsgemeinschaft, and Helmholtz-Gemeinschaft Deutscher Forschungszentren, Germany; the General Secretariat for Research and Technology, Greece; the National Scientific Research Foundation, and National Office for Research and Technology, Hungary; the Department of Atomic Energy and the Department of Science and Technology, India; the Institute for Studies in Theoretical Physics and Mathematics, Iran; the Science Foundation, Ireland; the Istituto Nazionale di Fisica Nucleare, Italy; the Korean Ministry of Education, Science and Technology and the World Class University program of NRF, Korea; the Lithuanian Academy of Sciences; the Mexican Funding Agencies (CINVESTAV, CONACYT, SEP, and UASLP-FAI); the Ministry of Science and Innovation, New Zealand; the Pakistan Atomic Energy Commission; the State Commission for Scientific Research, Poland; the Funda\c{c}\~ao para a Ci\^encia e a Tecnologia, Portugal; JINR (Armenia, Belarus, Georgia, Ukraine, Uzbekistan); the Ministry of Science and Technologies of the Russian Federation, the Russian Ministry of Atomic Energy and the Russian Foundation for Basic Research; the Ministry of Science and Technological Development of Serbia; the Ministerio de Ciencia e Innovaci\'on, and Programa Consolider-Ingenio 2010, Spain; the Swiss Funding Agencies (ETH Board, ETH Zurich, PSI, SNF, UniZH, Canton Zurich, and SER); the National Science Council, Taipei; the Scientific and Technical Research Council of Turkey, and Turkish Atomic Energy Authority; the Science and Technology Facilities Council, UK; the US Department of Energy, and the US National Science Foundation.

Individuals have received support from the Marie-Curie programme and the European Research Council (European Union); the Leventis Foundation; the A. P. Sloan Foundation; the Alexander von Humboldt Foundation; the Belgian Federal Science Policy Office; the Fonds pour la Formation \`a la Recherche dans l'Industrie et dans l'Agriculture (FRIA-Belgium); the Agentschap voor Innovatie door Wetenschap en Technologie (IWT-Belgium); and the Council of Science and Industrial Research, India.

\bibliography{auto_generated}

\cleardoublepage \appendix\section{The CMS Collaboration \label{app:collab}}\begin{sloppypar}\hyphenpenalty=5000\widowpenalty=500\clubpenalty=5000\textbf{Yerevan Physics Institute,  Yerevan,  Armenia}\\*[0pt]
S.~Chatrchyan, V.~Khachatryan, A.M.~Sirunyan, A.~Tumasyan
\vskip\cmsinstskip
\textbf{Institut f\"{u}r Hochenergiephysik der OeAW,  Wien,  Austria}\\*[0pt]
W.~Adam, T.~Bergauer, M.~Dragicevic, J.~Er\"{o}, C.~Fabjan, M.~Friedl, R.~Fr\"{u}hwirth, V.M.~Ghete, J.~Hammer\cmsAuthorMark{1}, M.~Hoch, N.~H\"{o}rmann, J.~Hrubec, M.~Jeitler, W.~Kiesenhofer, A.~Knapitsch, M.~Krammer, D.~Liko, I.~Mikulec, M.~Pernicka$^{\textrm{\dag}}$, B.~Rahbaran, C.~Rohringer, H.~Rohringer, R.~Sch\"{o}fbeck, J.~Strauss, A.~Taurok, F.~Teischinger, P.~Wagner, W.~Waltenberger, G.~Walzel, E.~Widl, C.-E.~Wulz
% ...
\vskip\cmsinstskip
\dag:~Deceased\\
1:~~Also at CERN, European Organization for Nuclear Research, Geneva, Switzerland\\
% ...

\end{sloppypar}
\end{document}